\documentstyle[12pt]{article}
\normalbaselines
\newcommand{\Ir}{Z\!\!\!Z}
\newcommand{\Ibb}[1]{ {\rm I\ifmmode\mkern
            -3.6mu\else\kern -.2em\fi#1}}
\newcommand{\ibb}[1]{\leavevmode\hbox{\kern.3em\vrule
     height 1.2ex depth -.3ex width .2pt\kern-.3em\rm#1}}

\newcommand{\Cx}{{\ibb C}}
\newcommand{\Rl}{{\Ibb R}}
\newcommand{\n}{n_i\otimes n_k}
\newcommand{\bi}{b_{ik}}
\newcommand{\A}{\cal A}
\newcommand{\U}{{\cal U}}
\newcommand{\C}{C_0\hat{\otimes}C_0}
\newcommand{\nn}{(n_i\otimes n_k-n_k\otimes n_i)}
\newcommand{\na}{n_i\wedge n_k}

\begin{document}
\begin{center}
\vspace*{1.0cm}

{\LARGE{\bf Discrete Mathematics and Physics on the Planck-Scale}}

\vskip 1.5cm

{\large {\bf Manfred Requardt }}

\vskip 0.5 cm

Institut f\"ur Theoretische Physik \\
Universit\"at G\"ottingen \\
Bunsenstrasse 9 \\
37073 G\"ottingen \quad Germany

\end{center}

\vspace{1 cm}

\begin{abstract}
Starting from the hypothesis that both physics, in
particular space-time and the physical vacuum, and the corresponding
mathematics are discrete on the Planck scale we develop a certain
framework in form of a '{\it cellular network}' consisting of cells
interacting with each other via bonds. Both the internal states of
the cells and the "strength" of the bonds are assumed to be dynamical
variables. In section 3 the basis is laid for a version of '{\it
discrete
analysis}' which, starting from different, perhaps more physically
oriented principles, manages to make contact with the much more
abstract machinery of Connes et al. and may complement the latter
approach. In section 4 a, as far as we can see, new concept of
'{\it topological dimension}' in form of a '{\it degree of
connectivity}' for graphs, networks and the like is developed. It is
then indicated how
this '{\it dimension}', which for continuous structures or lattices
being
embedded in a continuous background agrees with the usual
notion of dimension, may change dynamically as a result of a '{\it
phase
transition like}' change in '{\it connectivity}' in the network. A
certain
speculative argument, along the lines of statistical mechanics, is
supplied in favor of the naturalness of dimension 4 of ordinary
(classical) space-time.
\end{abstract} \newpage
\section{Introduction}
\noindent There exists a certain suspicion in parts of the scientific
community that nature may be "discrete" on the Planck scale. The
point of view held by the majority is however, at least as far as we
can see, that quantum theory as we know it holds sway down to
arbitrarily small scales as an allembracing general principle, being
applied to a sequence of increasingly fine grained effective field
theories all the way down up to, say, string field theory. But even
on that fundamental level one starts from strings moving in a
continuous background. It is then argued that "discreteness" enters
somehow through the backdoor via "quantisation".

The possibly most radical and heretical attempt, on the other
side, is it to try to generate both gravity and quantum theory as
secondary and
derived concepts (in fact merely two aspects) of one and the same
underlying
more primordial theory instead of simply trying to quantise gravity,
which is the canonical point of view (see e.g. \cite{1}).

This strategy implies more or less directly that -- as gravity is
closely linked with the dynamics of (continuous) space-time -- the
hypothetical underlying more fundamental theory is supposed to live
on a substratum which does not support from the outset something like
continuous topological or geometrical structures. In our view these
continuous structures should emerge as derived concepts via some sort
of coarse graining over a relatively large number of "discrete" more
elementary building blocks.

This program still leaves us with a lot of possibilities. For various
reasons, which may become more plausible in the course of the
investigation, we personally favor what we would like to call a
"cellular network" as a realisation of this substratum, the precise
definitions being given below. Without going into any details at the
moment some of our personal motivations are briefly the following:\\
i) These systems are in a natural way discrete, the local state space
at each site being usually finite or at least countable.\\
ii) Systems like these or their (probably better known) close
relatives, the "cellular automata", are known to be capable of
socalled "complex behavior", "pattern generation" and
"selforganisation" in general while the underlying dynamical laws are
frequently strikingly simple (a wellknown example being e.g.
Conway's "game of life").\\
Remark: A beautiful introduction into this fascinating field is e.g.
\cite{2}. As a shorter review one may take the contribution of
Wolfram (l.c.). More recent material can be found in the proceedings
of the Santa Fee Institute, e.g. the article of Kauffman in \cite{3},
who investigates slightly different systems ("switching nets").\\
iii) Some people suspect (as also we do) that physics may be
reducible at its very bottom to some sort of "information processing
system" (cf. e.g. \cite{4,5}). Evidently cellular automata and the
like are optimally adapted to this purpose.\\
iv) In "ordinary" field theory phenomena evolving in space-time are
typically described by forming a fibre bundle over space-time (being
locally homeomorphic to a product). In our view a picture like this
can only be an approximate one. It conveys the impression that
space-time is kind of an arena or stage being fundamentally different
from the various fields and phenomena which evolve and interact in
it. In our view these localised attributes, being encoded in the
various
field values, should rather be attributes of the -- in the
conventional picture hidden -- infinitesimal neighborhoods of
space-time points, more properly speaking, neighborhoods in a medium
in which space-time is immersed as a lower dimensional "submanifold".
To put it in a nutshell: We would prefer a medium in which what we
typically regard as irreducible space-time points have an internal
structure. To give a simple picture from an entirely different field:
take e.g. a classical gas, consider local pressure, temperature etc.
as collective coarse grained coordinates with respect to the
infinitesimal volume elements, regard then the microscopic degrees of
freedom of the particles in this small volume elements as the hidden
internal structure of the "points" given by the values of the above
collective coordinates (warning: this picture is of course not
completely correct as the correspondence between the values of
local pressure etc. and volume elements is usually not one-one).
It will turn out that a discrete structure as alluded to above is a
nice playground for modelling such features.\\[0.5cm]
Remark: Evidently there are close ties between what we have said in
iv) and certain foundational investigations in pure mathematics
concerning the problem of the 'continuum', a catchword being e.g.
"non-standard analysis".\vspace{0.5cm}

A lot more could be said as to the general physical motivations and a
lot more literature could be mentioned as e.g. the work of
Finkelstein and many others (see e.g. \cite{6,7}.  For further
references
cf. also the papers of Dimakis and M\"uller-Hoissen (\cite{8}).  Most
similar in spirit is in our view however the approach of 't Hooft
(\cite{9}).

\section{The Concept of the "Cellular Network"}
While our primary interest is in the analysis of various partly long
standing problems of current physics, which seem to beset physics
many orders away from the Planck regime, we nevertheless claim that
the understanding of the processes going on in the cellular network
at Planck level will provide us with strong clues concerning the
phenomena occurring in the "daylight" of
"middle-energy-quantum-physics". In fact, as Planck scale physics is
-- possibly for all times -- beyond the reach of experimental
confirmation, this sort of serious speculation has to be taken as a
substitution for experiments.

To mention some of these urgent problems of present day physics:\\
i) The unification of quantum theory and gravitation in general,\\
ii) the emergence of the universe, of space-time from "nothing" and
its very early period of existence,\\
iii) the mystery of the seeming vanishing of the 'cosmological
constant', which, in our view, is intimately related to the correct
understanding of the nature of vacuum fluctuations,\\
iv) the primordial nature of the "Higgs mechanism",\\
v) causality in quantum physics,\\
vi) 'potential' versus 'actual' existence in the quantum world and
the quantum mechanical measurement problem.

Some of these topics have been adressed by us recently in a somewhat
tentative way, based partly on the assumption that nature behaves or
can be imitated as a cellular network at its very bottom (\cite{10}).
The analysis was however hampered by the fact that the mathematical
and technical details of the underlying discrete model were at that
time not appropriately  developed to a sufficiently high degree.
Therefore we will concentrate in the following mainly on establishing
the necessary (mostly mathematical) prerequisites on which the
subsequent physical investigations can be safely based.

This is the more so necessary because one of our central hypotheses
is that most of the hierarchical structure and fundamental building
blocks of modern physics come into being via a sequence of {\it
unfolding
phase transitions} in this cellular medium. As far as we can see, the
study of phase transitions in cellular networks is not yet very far
developed, which is understandable given the extreme complexity of
the whole field. Therefore a good deal of work should be, to begin
with, devoted to a qualitative understanding of this intricate
subject.\\[0.5cm]
{\bf 2.1 Definition(Cellular Automaton)}: A cellular automaton
consists typically of a fixed regular array of cells \{$C_i$\}
sitting on the nodes \{$n_i$\} of a regular lattice like, say, $\Ir^d$
for some d.  Each of the cells is characterized by its internal state
$s_i$ which can vary over a certain (typically finite) set $\cal S$
which is usually chosen to be the same for all lattice sites.

Evolution or dynamics take place in discrete steps $\tau$ and is
given by a certain specific 'local law' $LL$:
\begin{equation}s_i(t+\tau)=ll(\{s'_j(t)\}\;\;\underline{S}(t+\tau)=LL
(\underline{S}(t)) \end{equation}
where t denotes a certain "{\it clock time}" (not necessarily
physical time),
$\tau$ the elementary clock time interval, \{$s'_j$\} the
internal states of the nodes of a certain local neighborhood of the
cell $C_i$, {\it ll} a map:
\begin{equation}ll: {\cal S}^n \to \cal S
\end{equation}
with n the number of neighbors occurring in (1), \underline{S}(t) the
global state at "time" t, {\it LL} the corresponding global map
acting on
the total state space $X:=\{\underline{S}\}$. {\it LL} is called {\it
reversible} if it is a bijective map of X onto itself. \vspace{0.5cm}

Cellular automata of this type behave generically already very
complicated (see \cite{2}). But nevertheless we suspect they are
still not complicated enough in order to perform the specific type of
complex behavior we want them to do. For one, they are in our view
too regular and rigid. For another, the occurring regular lattices
inherit quasi automatically such a physically important notion like
'{\it dimension}' from the underlying embedding space.

Our intuition is however exactly the other way round. We want to
generate something like dimension (among other topological notions)
via a dynamical process (of phase transition type) from a more
primordial underlying model which, at least initially, is lacking
such characteristic properties and features.

There exist a couple of further, perhaps subjective, motivations
which will perhaps become more apparent in the following and which
result in the choice of the following primordial model
system:\\[0.5cm]
{\bf 2.2 Definition(Cellular Network)}: In the following we will
mainly deal with the kind of system defined below:\\
i) "Geometrically" it is a {\it graph}, i.e. it consists of nodes
\{$n_i$\} and bonds \{$\bi$\} where pictorially the bond $\bi$
connects the nodes $n_i$ and $n_k$ with $n_i\neq n_k$ implied (there
are graphs where this is not so), furthermore, to each pair of nodes
there exists at most one bond connecting them.

The graph is assumed to be {\it connected}, i.e. two arbitrary nodes
can be connected by a sequence of consecutive bonds, and {\it
regular}, that is it looks locally the same everywhere.
Mathematically this means that the number of bonds being incident
with a given node is the same over the graph ({\it order of a node}).
We call the nodes which can be reached from a given node by making
one step the {\it 1-order-neighborhood} ${\cal U}_1$ and by not more
than n steps ${\cal U}_n$.\\
ii) On the graph we implant a dynamics in the following way:\\[0.5cm]
{\bf 2.3 Definition(Dynamics)}: As for a cellular automaton each node
$n_i$ can be in a number of internal states $s_i\in \cal S$. Each
bond $\bi$ carries a corresponding bond state $J_{ik}\in\cal J$. Then
we assume:
\begin{equation} s_i(t+\tau)=ll_s(\{s'_k(t)\},\{J'_{kl}(t)\})
\end{equation}
\begin{equation} J_{ik}(t+\tau)=ll_J(\{s'_l(t)\},\{J'{lm}(t)\})
\end{equation}
\begin{equation}
(\underline{S},\underline{J})(t+\tau)=LL((\underline{S},\underline{J}(
t))
\end{equation}
where $ll_s$, $ll_J$ are two mappings (being the same all over the
graph) from the state space of a local
neighborhood of a given fixed node or bond to $\cal S, J$, yielding
the
updated values of $s_i$ and $\bi$.\\[0.5cm]
Remarks: i) The theory of graphs is developed in e.g. \cite{11,12}.\\
ii) Synonyma for 'node' and 'bond' are e.g. 'site' and 'link'.\\
iii) It may be possible under certain circumstances to replace or
rather emulate  a cellular network of the above kind by some sort of
extended cellular automaton (e.g. by replacing the bonds by
additional sites). The description will then however become quite
cumbersome and involved. \vspace{0.5cm}

What is the physical philosophy behind this picture? We assume the
primordial substratum from which the physical universe is expected to
emerge via a selforganisation process to be devoid of most of the
characteristics we are usually accustomed to attribute to something
like a manifold or a topological space. What we are prepared to admit
is some kind of "{\it pregeometry}" consisting in this model under
discussion of an irregular array of elementary grains and "direct
interactions" between them, more specifically, between the members of
the various local neighborhoods.

It is an essential ingredient of our approach that the strength of
these direct interactions is of a dynamical nature and allowed to
vary. In particular it can happen that two nodes or a whole cluster
of nodes start to interact very strongly in the course of the
evolution and that this type of {\it collective behavior} persists
for a long time or forever (becomes {\it locked in}) or, on the other
extreme, that the interaction between certain nodes becomes weak or
vanishes.\\[0.5cm]
Remark: Note that -- in contrast to e.g. lattice field theory -- for
the time being the socalled '{\it clock time}' t is not standing on
the same footing as, say, potential coordinates in the network (e.g.
curves of bonds). We suppose anyhow that socalled '{\it physical
time}' will emerge as sort of a secondary collective variable in the
network, i.e. being different from the clock time (while being of
course functionally related to it).

In our view this is consistent with the spirit of relativity. What
Einstein was really teaching us is that there is a (dynamical)
interdependence between what we experience as space and time, not
that they are absolutely identical! \vspace{0.5cm}

As can be seen from the definition of the cellular network it
separates quite naturally into two parts of a different mathematical
and physical nature. The first one comprises part i) of definition
2.2, the second one part ii) and definition 2.3. The first one is
more static and "geometric" in character, the latter one conveys a
more dynamical and topological flavor as we shall see in the
following.

We begin in section 3 with a representation of what may be called
discrete analysis on graphs and networks. This is followed in section
4 by an investigation of certain dynamical processes in networks of
the defined type which resemble phase transitions and may induce {\it
dimensional change}. Most importantly we develop a physically
appropriate concept of'{\it dimension}' for such irregular discrete
structures which may be of importance in a wider context.
\section{Discrete Analysis on Networks}
At first glance one would surmise that as an effect of discreteness
something like a network will lack sufficient structure for such a
discipline to exist, but this is not so. Quite the contrary, there
are intimate and subtle relations to various recent branches of pure
mathematics as e.g. '{\it cyclic (co)homology}', '{\it noncommutative
de Rham complexes}', '{\it noncommutative geometry}' in general and
the like (see e.g. \cite{13}-\cite{16}).

The general aim of these recent developements is it to generate
something like a geometrical and differentiable structure within
certain mathematical contexts which traditionally are not considered
to support such structures.  Particularly simple examples are
discrete sets of, say, points, e.g. lattices. In a series of papers
Dimakis and M\"uller-Hoissen have applied the general abstract
machinery to models like these, having a possible bearing to, say,
lattice field theory etc. (see e.g. \cite{8} and further references
there).

The fundamental object in these approaches is typically the socalled
'{\it universal differential algebra}' or '{\it differential
envelope}' which can be canonically constructed over any associative
algebra and which is considered to be a generalisation or surrogate
(depending on the point of view) of a differential structure in the
ordinary cases.

As this notion may already indicate this scheme, paying tribute to its
universality and generality, is sometimes relatively far away from
the concrete physical models one is perhaps having in mind. In the
case of networks, for example, the inevitable starting point is the
'{\it maximally connected}' network or graph (also called a '{\it
complete graph}' or in algebraic topology a '{\it simplex}'), i.e. any
two nodes are directly connected by a bond.

As a consequence, the construction is lacking, at least initially,
something which is of tantamount importance in physical models, i.e.
a natural and physically motivated neighborhood structure. Typically
the interesting physical models are relatively lowly connected, which
implies that they usually exhibit a pronounced feeling of what is near
by or far away on the network.

One can of course pull this general structure down to the level of
the models one may have in mind by imposing '{\it relations}' between
various classes of '{\it differential forms}' but anyway, given a
concrete model this approach is relatively abstract and perhaps not
the most transparent and direct one. Furthermore, as it stresses more
the global algebraic relations, it does not naturally contain
from the beginning something eminently geometrical like e.g. '{\it
derivations at a
point}' which, on the other side, are fundamental in ordinary
differential topology.

There are other personal reasons to undertake to complement this
elegant but more algebraic framework by an approach which carries, at
least in our view, a more physical/geometric flavor and which is in
some sense oriented "bottom up" instead of "top down".

We begin with the introduction of some useful concepts borrowed from
algebraic topology and also known from graph theory (as to this we
recommend the beautiful book of Lefschetz, \cite{17}.

At first we have to give the graph an '{\it orientation}':\\[0.5cm]
{\bf 3.1 Definition(Orientation)}: With the notions defined in
definition 2.2 we say the bond $\bi$ points from node $n_i$ to node
$n_k$, the bond $b_{ki}$ from $n_k$ to $n_i$. We call $n_i$, $n_k$
initial and terminal node of $\bi$ respectively. We assume the up to
now formal relation:
\begin{equation} \bi=-b_{ki} \end{equation}
Remark: Note that orientation in the above (mathematical) sense is
different from what is understood in many applications as '{\it
directed bond}' in a network (as e.g. in typical "Kauffman nets",
\cite{3}). There a directed bond can typically "transport", say, a
message only in one given fixed direction. That is, nets of this type
behave, in physical terms, pronouncedly anisotropic locally. The
defintion 3.1, on the other side, is rather implementing something
like the orientation of curves.\\[0.5cm]
{\bf 3.2 Definition(Chain Complexes)}: We introduce, to begin with,
the two vector spaces $C_0$, $C_1$ whose elements, {\it zero- and
one-chains} are defined by up to now formal expressions
\begin{equation} \underline{c}_0 :=\sum f_in_i\quad \underline{c}_1
:=\sum g_{ik}b_{ik} \end{equation} \\[0.5cm]
where the $f_i$'s and $g_{ik}$'s range over a certain given field or
ring, of in the simplest cases numbers (i.e.$\Ir$,$\Rl$,$\Cx$), the
$n_i$'s and $g_{ik}$'s serve as generators of a free abelian
group.\\[0.5cm]
Remarks:i) Evidently one could in a next step choose much more general
spaces like, say, groups or manifolds.\\
ii) Furthermore, for the time being, the $f_i$'s and $g_{ik}$'s
should not be confused with the $s_i$'s and $J_{ik}$'s introduced in
section 2. The $f_i$'s and $g_{ik}$'s are e.g. allowed to vanish
outside a certain given cluster of nodes in various calculations or,
put differently, it may be convenient to deal only with certain
subgraphs.\\
iii) The spaces $C_0$, $C_1$ are in fact only the first two members
of a whole sequence of spaces.\\[0.5cm]
{\bf 3.3 Definition (Boundary)}: we now define a {\it boundary
operator} by
\begin{equation} \delta \bi:=n_k-n_i \end{equation}
which by linearity induces a linear map from $C_1$ to $C_0$:
\begin{equation} \delta: C_1\ni \sum g_{ik}\bi\to \sum
g_{ik}(n_k-n_i)\in C_0 \end{equation}
The kernel, $Z_1$ of this map, the 1-chains without '{\it boundary}',
consist of the '{\it 1-cycles}'. A typical example is a '{\it loop}',
i.e. a sequence of bonds, $\sum_{\nu}b_{i_{\nu}k_{\nu}}$ s.t.
$k_{\nu}=i_{\nu+1}$ and $k_n=i_1$. (However not every cycle is a
loop!).\\[0.5cm]
{\bf 3.4 Definition(Coboundary)}: Analogously we can define the
coboundary
operator as a map from $C_0$ to $C_1$:
\begin{equation} dn_i:=\sum_k \bi \end{equation}
where the sum extends over all bonds having $n_i$ as initial node,
and by linearity:
\begin{equation} d: \sum_i f_in_i \to \sum_i f_i(\sum_k \bi)
\end{equation}
Remark: Evidently one could in (8) and various other formulas choose
slightly different conventions, i.e. define $dn_i:= \sum_k b_{ki}$
etc. \vspace{0.5cm}

We will now show that these two operations, well known in algebraic
topology, can be fruitfully employed to create something like a
discrete calculus. Evidently, the 0-chains can as well be considered
as
functions over the set of nodes; in this case we abbreviate them by
f,g etc. chosen from  a certain subclass of 0-chains ${\cal A}
\subset
C_0$ (e.g. of '{\it finite support}', $L^1$, $L^2\ldots$). $\cal A$ is
trivially a module over itself (pointwise multiplication) freely
generated by the nodes $\{n_i\}$ which can be identified with the
'{\it elementary functions}' $e_i:=1\cdot n_i$.\\[0.5cm]
{\bf 3.5 Definition ("Quasi"Derivation)}: On a suitable class of node
functions $\cal A$ we call d a quasiderivation and df a differential.
That this is meaningful will become apparent from a rearrangement of
(11):

With $f_i\bi$ there occurs always $f_kb_{ki}=-f_k\bi$ on the rhs of
(11), hence we have:
\\[0.5cm]
{\bf 3.6 Observation}: \begin{equation} df=(\sum_i
f_in_i)=1/2\cdot\sum_{ik}(f_k-f_i)\bi \end{equation} \vspace{0.5cm}

We have still to show to what extent the operation d defined above
has the properties we are expecting from an (exterior) derivation.
The really crucial property in the continuum case is the (graded)
Leibniz rule. This is in fact a subtle and interesting point. To see
this we make a short aside about how discrete differentiation is
usually expected to work.

Take the following definition:\\[0.5cm]
{\bf 3.7 Definition (Partial Forward Derivative and Partial
Differential at Node (i))}:
\begin{equation} \nabla_{ik}f(i):=f(k)-f(i) \end{equation}
where $n_i,n_k$ are '{\it nearest-neighbor-nodes}', i.e. being
connected by a bond $\bi$.\\[0.5cm]
{\bf 3.8 Corollary}:
\begin{eqnarray}
\nabla_{ik}(f\cdot g)(i) & = & (f\cdot g)(k)-(f\cdot g)(i)
\nonumber\\
                         & = & \nabla_{ik}f(i)\cdot
g(i)+f(k)\cdot\nabla_{ik}g(i)\\
                         & = &
\nabla_{ik}f(i)g(i)+f(i)\nabla_{ik}g(i)+\nabla_{ik}f(i)\nabla_{ik}g(i)
\end{eqnarray}
In other words the "derivation" $\nabla$ does {\bf not} obey the
Leibniz rule. In fact, application of $\nabla$ to, say, higher powers
of f becomes increasingly cumbersome (nevertheless there is a certain
systematic in it). One gets for example:
\begin{equation} \nabla_{ik}f^{(n)}(i)=\nabla_{ik}f(i)\cdot
\{f^{(n-1)}(k)+f^{(n-2)}(k)f(i)+\ldots+f(k)f^{(n-2)}(i)+f^{(n-1)}(i)\}
\end{equation}

Due to the discreteness of the formalism and, as a consequence, the
inevitable bilocality of the derivative there is no chance to get
something as a true Leibniz rule on this level. (That this is
impossible has also been stressed from a different point of view in
e.g. example 2.1.1 of \cite{14}).

In some sense it is considered to be one of the merits of the
abstract algebraic framework (mentioned at the beginning of this
section) that a graded Leibniz rule holds in that generalized case.
The concrete network model under investigation offers a good
opportunity to test the practical usefulness of concepts like these.

To write down something like a Leibniz rule an important structural
element is still missing, i.e. the multiplication of node functions
from, say, some $\cal A$ with the members of $C_1$, in other words a
'{\it module structure over \cal A}. One could try to make the
following definition:
\begin{equation} f\cdot\bi:=f(i)\cdot\bi\quad \bi\cdot
f:=f(k)\cdot\bi \end{equation}
and extend this by linearity.

Unfortunately this "definition" does not respect the relation
$\bi=-b_{ki}$. We have in fact:
\begin{equation} f(i)\bi=f\cdot\bi=-f\cdot b_{ki}=-f(k)b_{ki}=f(k)\bi
\end{equation}
which is wrong in general for non-constant f!

Evidently the problem arises from our geometrical intuition which
results in the natural condition $\bi=-b_{ki}$, a relation we however
want to stick to. On the other side we can extend or embed our
formalism in a way which looses the immediate contact with
geometrical evidence but grants us with some additional mathematical
structure. (This is in fact common mathematical practice).

We can define another relation between nodes, calling two nodes
related if they are connected by a bond with a built-in direction
from the one to the other. We write this in form of a {\it tensor
product} structure. In the general tensor product $C_0\otimes C_0$ we
consider only the subspace $\C$ spanned by the elements $\n$ with
$n_i,n_k$ connected by a bond and consider $\n$ to be unrelated to
$n_k\otimes n_i$, i.e. they are assumed to be linearly independent
basis elements.\\[0.5cm]
{\bf 3.9 Observation}: There exists an isomorphic embedding of $C_1$
onto
the subspace generated by the antisymmetric elements in $\C$, i.e:
\begin{equation} \bi\to 1/2\cdot\nn=:\na \end{equation}
generate an isomorphism by linearity between $C_1$ and the
corresponding subspace $C_0\wedge C_0 \subset \C$.\\[0.5cm]
Proof: Both $\bi$ and $\na$ are linearly independent in there
respective vector spaces. \vspace{0.5cm}

In contrast to $C_1$ the larger $\C$ now supports a non-trivial
bimodule structure:\\[0.5cm]
{\bf 3.10 Observation/Defintion (Bimodule)}: We can now define
\begin{eqnarray} f\cdot (\n) & := & f(i)(\n)\\
          (\n)\cdot f        & := & f(k)(\n)
\end{eqnarray}
and extend this by linearity to the whole $\C$, making it into a
bimodule over some $\cal A \in C_0$.\\[0.5cm]
Remark: Equivalently one could replace $\n$ by $e_i\otimes e_k$, the
corresponding elementary functions.\\[0.5cm]
{\bf 3.11 Corollary}: As a module over $\cal A$, $\C$ is generated by
$C_0\wedge C_0$.\\[0.5cm]
Proof: It suffices to show that every $\n$ can be generated this way.
\begin{equation} n_i\cdot\nn=\n \end{equation}
as $n_i\cdot n_k=0$ for $i\neq k$. \vspace{0.5cm}

With the $\bi$ so embedded in a larger space and identified with
\begin{equation}
1/2\cdot\nn=\na
\end{equation}
we are in the position to derive a graded Leibniz
rule on the module (algebra) $\A$. Due to linearity and the structure
of the respective spaces it suffices to show this for products of
elementary functions $e_i=n_i$. We in fact have:\\[0.5cm]
($i\neq k$ not nearest neighbors):\begin{equation}d(n_i\cdot
n_k)=0,\,dn_i\cdot n_k=n_i\cdot dn_k=0 \end{equation}
($i\neq k$ nearest neighbors):\begin{equation}d(n_i\cdot
n_k)=d(0)=0\;\;\mbox{and} \end{equation}
\begin{eqnarray}dn_i\cdot n_k+n_i\cdot dn_k & = &
(\sum_{k'}b_{ik'})\cdot n_k+n_i\cdot(\sum_{i'}b_{ki'})\\ & = &
\bi\cdot n_k+n_i\cdot b_{ki}\\
& = & 1/2\{\nn n_k+n_i(n_k\otimes n_i-n_i\otimes
n_k)\}\\ & = & 1/2\{n_i\otimes n_k-n_i\otimes n_k\}=0 \end{eqnarray}
$(i=k)$:
\begin{equation} d(n_i^2)=d(n_i)=\sum_k\bi\;\;\mbox{and}
\end{equation}
\begin{eqnarray} dn_i\cdot n_i+n_i\cdot dn_i & = &
(\sum_k\bi)\,n_i+n_i(\sum_k\bi)\\ & = & 1/2\sum_k\nn=\sum_k\bi=dn_i
\end{eqnarray}
{\bf 3.12 Conclusion}: In the bimodule $\C$ generated by the elements
$\bi$
over $\A$ the map d fulfills the Leibniz rule, i.e:
\begin{equation} d(f\cdot g)=df\cdot g+f\cdot dg \end{equation}

{}From the above we see also that functions, i.e. elements from $\A$
and bonds or differentials of functions do no longer commute (more
specifically, the two possible ways of imposing a module structure
could be considered this way). We have for example:\\[0.5cm]
{\bf 3.13 Commutation Relations}:\\
($i\neq k$ not nearest neighbors): \begin{equation} n_i\cdot
dn_k=dn_k\cdot n_i=0 \end{equation}
($i\neq k$ nearest neighbors). \begin{eqnarray} n_i\cdot
dn_k+dn_k\cdot n_i & = & 1/2\sum_{i'}\{n_i(n_k\otimes
n_{i'}-n_{i'}\otimes n_k)\\ & + & (n_k\otimes
n_{i'}-n_{i'}\otimes n_k)\,n_i\}\\ & = & -1/2\nn=-\bi \end{eqnarray}
($i=k$): \begin{equation} n_i\cdot dn_i+dn_i\cdot n_i=\sum_k\bi=dn_i
\end{equation}

Making contact with local differential topology on manifolds we can
now formulate the following concepts:\\[0.5cm]
{\bf 3.14 Definition ((Co)Tangential Space)}:\\i) We call the space
spanned by the $\nabla_{ik}$ at node $n_i$ the tangential space
$T_i$.\\ ii) Correspondingly the space spanned by $\bi$ at node $n_i$
is called the cotangential space $T_i^{\ast}$.\\ We can now consider
the $b_{ik}$'s as linear forms over $T_i$ via:
\begin{equation} <\bi|\nabla_{ij}>=\delta_{kj} \end{equation}

Another important relation we want to mention is the followng:
$\delta\,d\,f$ is a map from $C_0\to C_0$ and reads in
detail:\\[0.5cm]
{\bf 3.15 Observation (Laplacian)}: \begin{equation}
\delta\,d\,f=-\sum_i(\sum_k f(k)-n\cdot f(i))\,n_i=:-\Delta\,f
\end{equation}
with n the number of nearest neighbors of $n_i$.\\[0.5cm]
Proof: \begin{eqnarray} \delta\,d\,f & = &
1/2\sum_{ik}(f(k)-f(i))(n_k-n_i)\\ & = &
1/2\sum_{ik}(f(k)\,n_k+f(i)\,n_i-f(i)\,n_k-f(k)\,n_i)\\ & = &
-\sum_i(\sum_kf(k)-n\cdot f(i))\,n_i \end{eqnarray}

Having now established the first steps in setting up this particular
version of discrete calculus one can proceed in various directions.
First, one can develop a discrete Lagrangian variational calculus ,
derive Euler-Lagrange-equations and Noetherian theorems and the like
and compare our approach with other existing schemes in discrete
mathematics.

Second, one can continue the above line of reasoning and proceed to
higher differential forms.\\[0.5cm]
{\bf 3.16 Definition/Observation}: Higher tensor products of
differential forms
at a node $n_i$ can be defined as {\it multilinear forms}:
\begin{equation} <b_{ik_1}\otimes\cdots\otimes
b_{ik_n}|(\nabla_{il_1},\cdots,\nabla_{il_n})>:=\delta_{k_1l_1}
\times\cdots{\times} \delta_{k_nl_n} \end{equation}
and linear extension.\vspace{0.5cm}

A comparison of our scheme with the ordinary approach, performed
within the
framework of the universal differential algebra, is, on the other
side, a subtle and delicate point and would lead us a little bit
astray at the
moment. The deeper reason is the following:

In contrast to the universal differential algebra mentioned above,
where every two nodes are connected by a bond, this is not so for our
'{\it reduced}' calculus. As a consequence certain operations are
straightforward to define in the former approach. However, descending
afterwards  to the lower-connected more realistic models is tedious
in general and not always particularly transparent. That is, this
method does not really save calculational efforts in typical concrete
cases (for a discussion of certain simple examples see \cite{8}).

The mathematical "triviality " of the differential envelope is
reflected by the trivialty of the corresponding '{\it (co)homology
groups}' of the maximally connected graph (simplex). This trivialty
is then broken by deleting graphs in the reduction process.

To mention a typical situation: Take e.g. the subgraphs of a graph G
consisting of, say, four nodes $n_i,n_k,n_l,n_m$
and all the bonds between them which occur in G.

In the case G being a simplex (i.e. non-reduced case) all these
subgraphs are geometrically/topologically equivalent. An important
consequence of this is that the four nodes can be connected by a
'{\it path}', i.e. a sequence of consecutive bonds with each node
being passed
only once, the effect being that one can naturally define a
multiplication in this scheme via '{\it concatenation}', e.g:
\begin{equation} (\n)\cdot(n_l\otimes
n_m)=n_i\otimes n_k\cdot n_l\otimes n_m \end{equation}
with $n_k\cdot n_l=\delta_{kl}\cdot n_k$. Correspondingly, all the
higher differentials can be generated by the 1-forms via
concatenation. The reason why it is sufficient to concatenate only at
the extreme left and right of a '{\it word}' stems exactly from the
simplex-character of each subgraph.

In typical reduced cases all this is no longer the case; the
combinatorical topology becomes non-trivial. To give an example: Take
as G a regular graph with the order of the nodes (number of incident
bonds) $n=3$. In the analogous case of 4-node-subgraphs there exists
now a kind of subgraph the nodes of which cannot be concatenated in
the above way. Take e.g. the subgraph with bonds existing only
between, say, $n_1\,n_4$, $n_2\,n_4$, $n_3\,n_4$. I.e., one has the
1-forms $b_{14},\;b_{24},\;b_{34}$ or:
\begin{equation} n_1\otimes n_4,\;n_2\otimes n_4,\;n_3\otimes n_4
\end{equation}
but there is no obvious way to generate the corresponding reduced
subgraph by concatenating them sequentially(!). The loophole is that
one has to define multiplication differently (this can in fact be
done and all the higher differentials generated that way  without
employing the universal differential envelope).

Nevertheless, the situation is much more involved in the more
realistic cases. As this highly interesting feature, which we have
not yet found discussed in this particular context in the literature
known to us, deserves a more careful analysis of its own we prefer to
present it elsewhere and proceed in the next section with the
developement
of a concept of '{\it dimension}' in networks and graphs which
reflects
the '{\it degree of connectivity}' and has a bearing on physical
concepts
like '{\it interaction}' and '{\it phase transitions}'.

\section{Intrinsic Dimension in Networks and other Discrete Systems}
There exist a variety of concepts in modern mathematics which
generalize the notion of '{\it dimension}' one is accustomed to in
e.g. differential topology or linear algebra. In fact, '{\it
topological dimension}' is a notion which seems to be even closer to
the underlying intuition (cf. e.g. \cite{18}).

Apart from the purely mathematical concept there is also a physical
aspect of something like dimension which has e.g. pronounced effects
on the behavior of, say, many-body-systems, especially their
microscopic dynamics and, most notably, their possible '{\it phase
transitions}'.

But even in the case of e.g. lattice systems they are usually
considered as embedded in an underlying continuous background space
(typically euclidean) which supplies the concept of ordinary
dimension so that the intrinsic dimension of the discrete array itself
does usually not openly enter the considerations.

Anyway, it is worthwhile even in this relatively transparent
situations to have a closer look on where attributes of something
like dimension really come into the physical play. Properties of
models of, say, statitical mechanics  are almost solely derived from
the structure of the microscopic interactions of their constituents.
This is more or less the only place where dimensional aspects enter
the calculations.

Naive reasoning might suggest that it is the number of nearest
neighbors (in e.g. lattice systems) which reflects in an obvious way
the dimension of the underlying space and influences via that way the
dynamics of the system. However, this surmise, as we will show in the
following, does not reflect the crucial point which is considerably
more subtle.

This holds the more so for systems which cannot be consided as being
embedded in a smooth regular background and hence do not get their
dimension from the embedding space. A case in point is our primordial
network in which in which Planck-scale-physics is assumed to tkae
place. In our approach it is in fact exactly the other way round:
Smooth space-time is assumed to emerge via a phase transition and
after some '{\it coarse graining}' from this more fundamental
structure.\\[0.5cm]
{\bf 4.1 Problem}: Formulate an intrinsic notion of dimension for
model theories without making recourse to the dimension of some
embedding space.\vspace{0.5cm}

In a first step we will show that graphs and networks as introduced
in the preceding sections have a natural metric structure. We have
already introduced a certain neighborhood structure in a graph with
the help of the minimal number of consecutive bonds connecting two
given nodes.

In a connected graph any two nodes can be connected by a sequence of
bonds. Without loss of generality one can restrict oneself to '{\it
paths}'. One can then define the length of a path (or sequence of
bonds) by the number l of consecutive bonds making up the
path.\\
{\bf 4.2 Observation/Definition}: Among the paths connecting two
arbitrary nodes there exists at least one with minimal length which
we denote by $d(n_i,n_k)$. This d has the properties of a '{\it
metric}', i.e:
\begin{eqnarray} d(n_i,n_i) & = & 0\\ d(n_i,n_k) & = &
d(n_k,n_i)  \\d(n_i,n_l) & \leq & d(n_i,n_k)+d(n_k,n_l) \end{eqnarray}
(The proof is more or less evident).\\[0.5cm]
{\bf 4.3 Corollary}: With the help of the metric one gets a natural
neighborhood structure around any given node, where ${\cal U}_m(n_i)$
comprises all the nodes with $d(n_i,n_k)\leq m$, $\partial{\cal
U}_m(n_i)$
the nodes with $d(n_i,n_k)=m$. \vspace{0.5cm}

With the help of the above neighborhood structure we can now develop
the concept of an intrinsic dimension on graphs and networks. To this
end one has at first to realize what property really matters
physically (e.g. dynamically) independently of the model or embedding
space. \\[0.5cm]
{\bf 4.4 Observation}: The crucial and characteristic property of,
say, a graph or network which may be associated with something like
dimension is the number of '{\it new nodes}' in ${\cal U}_{m+1}$
compared
to ${\cal U}_m$ for m sufficiently large or $m\to \infty$. The deeper
meaning of this quantity is that it measures the kind of '{\it
wiring}' or '{\it connectivity}' in the network and is therefore a
'{\it topological invariant}'.\\[0.5cm]
Remark: In the light of what we have learned in the preceding section
it is tempting to relate the number of bonds branching off a node,
i.e. the number of nearest neighbors or order of a node, to something
like dimension.

On the other side there exist quite a few different lattices with a
variety of number nearest neighbors  in, say, two- or three-
dimensional euclidean space. What however really matters in physics
is the embedding dimension of the lattice (e.g. with respect to phase
transitions) and only to a much lesser extent the number of nearest
neighbors.

In contrast to the latter property dimension reflects the degree of
connectivity and type of wiring in the network. \vspace{0.5cm}

In many cases one expects the number of nodes in ${\cal U}_m$ to grow
like
some power D of m for increasing m. By the same token one expects the
number of new nodes after an additional step to increase proportional
to $m^{D-1}$. With $|\,\cdot\,|$ denoting number of nodes
we hence have:
\begin{equation} |{\cal U}_{m+1}|-|{\cal U}_m|=|\partial {\cal
U}_{m+1}|=f(m)
\end{equation}
with
\begin{equation} f(m)\sim m^{D-1} \end{equation}
for m large.\\[0.5cm]
{\bf 4.5 Definition}: The intrinsic dimension D of a regular
(infinite) graph is given by
\begin{equation} D-1:=\lim_{m\to \infty}(\ln f(m)/\ln m)\; \mbox{or}
\end{equation}
\begin{equation} D:=\lim_{m\to \infty}(\ln |{\cal U}_m|/\ln m)
\end{equation}

That this definition is reasonable can be seen by applying it to
ordinary cases like regular translation invariant lattices. \\[0.5cm]
{\bf 4.6 Observation} For regular lattices D coincides with the
dimension of the euclidean embedding space $D_E$.\\[0.5cm]
Proof: It is instructive to draw a picture of the consecutive series
of neighborhoods of a fixed node for e.g. a 2-dimensional Bravais
lattice. It is obvious and can also be proved that for m sufficiently
large the number of nodes in $\U_m$ goes like a power of m with the
exponent being the embedding dimension $D_E$ as the euclidean volume
of $\U_m$ grows with the same power.\\[0.5cm]
Remarks:i) For $\U_m$ to small the number of nodes may deviate from
an exact power law which in general becomes only correct for
sufficiently large m.
\\ii) The number of nearest neighbors, on the other side, does not(!)
influence the exponent but rather enters in the prefactor. In other
words, it influences $|\U_m|$ for m small but drops out
asymptotically by taking the logarithm. For an ordinary Bravais
lattice with $N_C$ the number of nodes in a unit cell one has
asymptotically:
\begin{equation} |\U_m|\sim N_C\cdot m^{D_E} \quad\mbox{and hence:}
\end{equation}
\begin{equation} D=\lim_{m\to\infty}(\ln(N_C\cdot m^{D_E})/\ln
m)=D_E+\lim_{m\to\infty}(N_C/\ln m)=D_E
\end{equation}
independently of $N_C$.\vspace{0.5cm}

Matters become much more interesting and subtle if one studies more
general graphs than simple lattices. Note that there exists a general
theorem showing that practically every graph can be embedded in
$\Rl^3$ and still quite a lot in $\Rl^2$ ('{\it planar graphs}').

The point is however that this embedding is in general not invariant
with respect to the euclidean metric. But something like an apriori
given euclidean length is unphysical for the models we are after
anyhow.
This result has the advantage that one can visualize many graphs
already in, say, $\Rl^2$ whereas their intrinsic dimension may be much
larger.

An extreme example is a '{\it tree graph}', i.e. a graph without
'{\it loops}'. In the following we study an infinite, regular tree
graph with node order 3, i.e. 3 bonds branching off each node. The
absence of loops means that the '{\it connectivity}' is extremely low
which results in an exceptionally high '{\it dimension}' as we will
see.

Starting from an arbitrary node we can construct the neighborhoods
$\U_m$ and count the number of nodes in $\U_m$ or $\partial\U_m$.
$\U_1$ contains 3 nodes which are linked with the reference node
$n_0$. There are 2 other bonds branching off each of these nodes.
Hence in $\partial\U_2=\U_2\backslash\U_1$ we have $3\cdot2$ nodes
and by induction:
\begin{equation} |\partial\U_{m+1}|=3\cdot2^m \end{equation}
which implies
\begin{equation} D-1:=\lim_{m\to\infty}(\ln|\partial\U_{m+1}|/\ln m)=
\lim_{m\to\infty}(m\cdot ln 2/\ln m+3/\ln m)=\infty \end{equation}
Hence we have: \\[0.5cm]
{\bf 4.7 Observation(Tree)}: The intrinsic dimension of an infinite
tree is $\infty$ and the number of new nodes grows exponentially like
some $n(n-1)^m$ with m (n being the node order).\\[0.5cm]
Remark: $D=\infty$ is mainly a result of the absence of loops(!), in
other words: there is exactly one path, connecting any two nodes.
This is usually not so in other graphs, e.g. lattices, where the
number of new nodes grows at a much slower pace (whereas the number
of nearest neighbors can nevertheless be large). This is due to the
existence of many loops s.t. many of the nodes which can be reached
from, say, a node of $\partial\U_m$ by one step are already
contained in $\U_m$ itself. \vspace{0.5cm}

We have seen that for, say, lattices the number of new nodes grows
like some fixed power of m while for, say, trees m occurs in the
exponent. The borderline can be found as follows:\\[0.5cm]
{\bf 4.8 Observation}: If for $m\to\infty$ the average number of new
nodes per node contained in $\partial\U_m$, i.e:
\begin{equation} |\U_{m+1}|/|\U_m|\geq 1+\varepsilon \end{equation}
for some $\varepsilon\geq 0$ then we have exponential growth, in other
words, the intrinsic dimension is $\infty$. \\[0.5cm]
Proof: If the above estimate holds for all $m\geq m_0$ we have by
induction:
\begin{equation} |\U_m|\geq |\U_{m_0}|\cdot (1+\varepsilon)^{m-m_0}
\end{equation}

Potential applications of this concept of intrinsic dimension are
manifold. Our main goal is it to develop a theory which explains how
our classical space-time and what we call the '{\it physical vacuum}'
has emerged from a more primordial and discrete background via some
sort of phase transition.

In this context we can also ask in what sense space-time dimension 4
is exceptional, i.e. whether it is merely an accident or whether
there is a cogent reason for it.

As the plan of this paper was mainly to introduce and develop the
necessary
conceptual tools and to pave the ground, the bulk of the investigation
in this particular direction shall be presented elsewhere as it is
part of a detailed analysis of the (statitical) dynamics on networks
as introduced above, their possible phase transitions,
selforganisation, emergence of patterns and the like.

In this paper, which is to have a more technical flavor, we will only
supply a speculative and heuristic argument in favor of space-
dimension
3. As we emphasized in section 2 also the bond states, modelling the
strength of local interactions between neighboring nodes, are in our
model theory dynamical variables. In extreme cases these couplings
may completely vanish or become extremely strong between certain
nodes.

It may now happen that in the course of evolution a local island (or
several of them) emerges as a fluctuation in a, on large scales,
unordered or lowly connected network (e.g. a tree-like structure)
where couplings between nodes are switched on which have been
uncoupled before or, on the other side, become very strong.

One important effect of this scenario (among others) is that there
may occur
now a lot of local loops in this island while the state outside is
more or less loopless. This may have the consequence that the
intrinsic dimension within this island may become substantially lower
than outside, say, finite as compared to (nearly) infinity.

If this '{\it nucleation center}' is both sufficiently large and its
local state '{\it dynamically favorable}' in a sense to be specified
(note that a concept like '{\it entropy}' or something like that
would be of use here) it may start to grow and trigger a global phase
transition.

As a result of this phase transition a relatively smooth and stable
submanifold (in the language of synergetics an '{\it order parameter
manifold}') may come into being which displays certain properties we
would attribute to space-time.

Under these premises we could now ask what is the probability for
such a specific and sufficiently large spontaneous fluctuation? As we
are at the moment talking about heuristics and qualitative behavior
we make the following thumb-rule-like assumptions:\\[0.5cm]
i) In the primordial network '{\it correlation lengths}' are
supposed to be extremely short (more or less nearest neighbor), i.e.
the strengths of the couplings are fluctuating more or less
independently.\\
ii) A large fluctuation of the above type implies that a substantial
fraction of the couplings in the island passes a certain threshold,
i.e. becomes sufficiently strong and cooperative. The probability per
individual bond for this to happen be p. Let L be the diameter of the
nucleation center, $const\cdot L^d$ the number of nodes or bonds in
this island for some d. The probability for such a fluctuation is
then roughly:
\begin{equation} W_d=const\cdot p^{(L^d)} \end{equation}
iii) We know from our experience with phase transitions that there
are favorable dimensions, i.e. the nucleation centers may fade away
if either they themselves are too small or the dimension of the
system is too small. Apart from certain non-generic models $d=3$ is
typically the threshold dimension.\\
iv) On the other side we can compare the relative probabilities for
the occurrence of sufficiently large spontaneous fluctuations for
various d's. One has:
\begin{equation} W_{d+1}/W_d\sim p^{(L^{d+1})}/p^{(L^d)}=p^{L^d(L-1)}
\end{equation}
Take e.g. $d=3,\,L=10,\,p=1/2$ one gets:
\begin{equation} W_4/W_3\sim 2^{-(9\cdot10^3)} \end{equation}
In other words, provided that this crude estimate has a grain of
truth in it, one may at least get a certain clue that space-dimension
3 is both the threshold dimension and, among the class of in principle
allowed dimensions (i.e. $d\geq3$) the one with the dominant
probability.


\begin{thebibliography}{99}
\bibitem{1}C.J.Isham: "Conceptual and Geometrical Problems in Quantum
Gravity", Lecture presented at the 1991 Schladning Winter School
\bibitem{2}Contributions in Physica 10D(1984), especially the review
by S.Wolfram
\bibitem{3}S.Kauffman: in "Complexity, Entropy and the Physics of
Information, SFI-Studies Vol.VIII p.151, Ed. W.H.Zurek,
Addison-Wesley 1990
\bibitem{4}K.Zuse: in Act.Leop. Vol.37/1 ("Informatik") 1971 p.133,\\
C.F.von Weizs\"acker: l.c. p.509\\
K.Zuse: Int.J.Th.Phys. 21(1982)589
\bibitem{5}R.Feynman: as quoted in D.Finkelstein: Phys.Rev.
184(1969)1261 or:\\ Int.J.Th.Phys. 21(1982)467; in fact most of the
numbers 3/4, 6/7, 12 are devoted to this topic.
\bibitem{6}D.Finkelstein: Int.J.Th.Phys. 28(1989)1081
\bibitem{7}L.Bombelli, J.Lee, D.Meyer, R.Sorkin: Phys.Rev.Lett.
59(1987)521
\bibitem{8}A.Dimakis, F.M\"uller-Hoissen: J.Math.Phys. 35(1994)6703,\\
F.M\"uller-Hoissen: "Physical Aspects of Differential Calculi on
Commutative Algebras", Karpacz Winter School 1994
\bibitem{9}G.'t Hooft: J.Stat.Phys. 53(1988)323,\\
Nucl.Phys. B342(1990)471
\bibitem{10}M.Requardt: Preprints G\"ottingen 1993/94
\bibitem{11}O.Ore:"Theory of Graphs", American Math. Soc., N.Y. 1962
\bibitem{12}K.Wagner: "Graphentheorie", Bibliographisches Inst.,
Mannheim 1970
\bibitem{13}A.Connes: "Non-Commutative Geometry", Acad.Pr., N.Y. 1994
\bibitem{14}J.Madore: "Non-Commutative Differential Geometry and its
Physical Applications", LPTHE Orsay 1994
\bibitem{15}P.Seibt: "Cyclic Homology of Algebras", World Scientific,
Singapore 1987
\bibitem{16}D.Kastler: "Cyclic Cohomology within the Differential
Envelope", Hermann, Paris 1988
\bibitem{17}S.Lefschetz: "Applications of Algebraic Topology",
Springer, N.Y. 1975
\bibitem{18}G.A.Edgar: "Measure, Topology, and Fractal Geometry",
Springer, N.Y. 1990,\\
K.Kuratowski: "Topology" Vol.1, Acad.Pr., N.Y. 1966
\end{thebibliography}
\end{document}